\documentstyle[12pt,epsfig]{article}

\def\beq{\begin{equation}}  
\def\eeq{\end{equation}}  
\def\beqa{\begin{eqnarray}}  
\def\eeqa{\end{eqnarray}}   
  
\begin{document}  
\input axodraw.sty
   
\begin{titlepage}  
  
\begin{flushright}
{\sc UMHEP-454}\\ [.2in]  
{\sc October 20, 1998}\\ [.5in]  
\end{flushright}  
  
\begin{center}  
   
{\LARGE
Form factor relations for heavy-to-light meson transitions:
tests of the Quark Model predictions.}\\
[.5in]
{\large Jo\~{a}o M. Soares}\\ [.1in]  
{\small
Department of Physics and Astronomy\\  
University of Massachusetts\\
Amherst, MA 01003-4525}\\ [.5in]  
\end{center}

{\normalsize\bf
Abstract}

In the amplitudes for the weak semi-leptonic decays of mesons, the ha\-dron\-ic
matrix elements are parametrized by form factors that describe the
non-perturba\-ti\-ve QCD effects and are a source of large theoretical 
uncertainties. In the case of heavy-to-light meson transitions, a Quark Model 
derivation leads to very general relations between those hadronic form factors,
such that only two remain that are independent --- one for 
pseudoscalar-to-pseudoscalar and one for pseudoscalar-to-vector meson 
transitions. Here, we investigate to what extent these form factor relations 
remain a good approximation, beyond the Quark Model.

In heavy-to-light pseudoscalar-to-vector meson transitions, a simple argument 
shows that the V--A structure of the weak interaction leads to a strong 
suppression of the helicity $\lambda = +1$ amplitude --- an effect that has
been confirmed experimentally by the CLEO Collaboration, with a full angular 
analysis of the $B \to K^\ast J/\psi$ decay. We show that the theoretical 
predictions, in terms of the hadronic form factors, can accommodate the 
suppression of the $\lambda = +1$ amplitude, only if the Quark Model relations
are verified. Moreover, the form factor relations also allow us to predict the
ratio of the two remaining helicity amplitudes, with $\lambda = 0$ and $-1$; 
here too, there is excellent agreement with the CLEO data for the $B \to K^\ast
J/\psi$ decay. In the future, similar experimental tests can be carried out, 
with a few advantages, using the semileptonic decay $B \to \rho l^- \overline
{\nu}_l$. 

The Quark Model relations can also be tested against the predictions for the
hadronic form factors, from the more powerful theoretical methods of Lattice 
QCD and Light Cone Sum-Rules. The excellent agreement points once again to the
validity of the form factor relations, beyond the Quark Model framework where
they were derived.

{\small
PACS: 13.20.He, 13.20.Fc, 13.25.Ft, 13.25.Hw.}   
  
\end{titlepage}

\section{Introduction}  
 
The study of the weak decays of quarks is hampered by the presence of the long
distance QCD effects that are responsible for the binding of the quarks into 
hadrons. These non-perturbative effects are hard to evaluate in a model 
independent way, and so tend to bring large uncertainties to the theoretical 
predictions. They appear in the matrix elements of the weak Hamiltonian 
operators, between the initial and final hadronic states:
\beq
\langle X(\vec{p^\prime})|\overline{q} \Gamma b|B(\vec{p})\rangle \ ,  
\hspace{.3in}
\Gamma \equiv \gamma^\mu,\; \gamma^\mu \gamma_5,\; 
i \sigma^{\mu\nu} (p - p^\prime)_\nu,\;
i \sigma^{\mu\nu} (p - p^\prime)_\nu \gamma_5 \ .
\label{10}
\eeq
We are interested in the case of heavy-to-light transitions, where $q \equiv 
u$, $d$ or $s$ and the final state meson $X$ is a light pseudoscalar $P \equiv
\pi, \dots$, or a light vector meson $V \equiv \rho, \dots$; the initial state
meson is a heavy pseudoscalar $B$ meson, although the results may also be 
valid, to some degree, for the case of the lighter $D$ meson. The hadronic 
matrix elements in Eq.~\ref{10} are parametrized in terms of Lorentz invariant
form factors as follows:
\beqa 
\langle P(\vec{p^\prime})|\overline{q}\gamma^\mu b|B(\vec{p})\rangle
&=& (p + p^\prime)^\mu f_1(q^2) \nonumber \\
& & + \frac{m_B^2 - m_P^2}{q^2} q^\mu \left[ f_0(q^2) - f_1(q^2) \right] \ ,
\label{20} 
\eeqa 
where $f_1(0) = f_0(0)$;
\beqa 
\langle P(\vec{p^\prime})| \overline{q} i \sigma^{\mu\nu} q_\nu b 
|B(\vec{p})\rangle &=& 
s(q^2) \left[ (p + p^\prime)^\mu q^2 - (m_B^2 - m_P^2) q^\mu \right] \ ; 
\label{30} \\
\langle V(\vec{p^\prime},\vec{\varepsilon})|\overline{q}\gamma^\mu b
|B(\vec{p})\rangle 
&=& \frac{-1}{m_B + m_V} 2 i \epsilon^{\mu\alpha\beta\gamma}  
\varepsilon_\alpha^\ast p_\beta^\prime p_\gamma V(q^2) \ ; 
\label{40} \\
\langle V(\vec{p^\prime},\vec{\varepsilon})|\overline{q}\gamma^\mu \gamma_5
b|B(\vec{p})\rangle
&=& (m_B + m_V) \varepsilon^{\mu\ast} A_1(q^2)  \nonumber \\ 
& & -\frac{\varepsilon^\ast.q}{m_B + m_V} (p+p^\prime)^\mu A_2(q^2)  
\nonumber \\ 
& & - 2 m_V \frac{\varepsilon^\ast.q}{q^2} q^\mu 
\left[ A_3(q^2) - A_0(q^2) \right] \ ,
\label{50} 
\eeqa
where $2 m_V A_3(q^2) \equiv  (m_B + m_V) A_1(q^2) 
- (m_B - m_V) A_2(q^2)$ and $A_0(0) = A_3(0)$;
\pagebreak
\beqa 
\langle V(\vec{p^\prime},\vec{\varepsilon})| \overline{q} i  
\sigma^{\mu\nu} q_\nu b|B(\vec{p})\rangle &=& 
i \epsilon^{\mu\alpha\beta\gamma}
\varepsilon_\alpha^\ast p_\beta^\prime p_\gamma F_1(q^2) \ ; 
\label{60} \\
\langle V(\vec{p^\prime},\vec{\varepsilon})| \overline{q} i  
\sigma^{\mu\nu} q_\nu \gamma_5 b|B(\vec{p})\rangle &=&
\left[ (m_B^2 - m^2_V) \varepsilon^{\mu\ast}  
- \varepsilon^\ast.q (p+p^\prime)^\mu \right] F_2(q^2) \nonumber \\ 
& & + \varepsilon^\ast.q \left[ q^\mu - \frac{q^2}{m_B^2 - m^2_V}  
(p+p^\prime)^\mu \right] F_3(q^2) \ , \nonumber \\
& &
\label{70}
\end{eqnarray} 
where $F_1(0) = 2 F_2(0)$. In all of the above, $q \equiv p-p^\prime$. There 
are three form factors --- $f_{0,1}$ and $s$ --- for a $B \to P$ transition, 
and seven form factors --- $V$, $A_{0,1,2}$ and $F_{1,2,3}$ --- for a $B \to 
V$ transition. These form factors contain the long distance QCD effects and 
are therefore poorly known; relations between them, that will hold under 
certain conditions or approximations, can then be very useful: they will 
reduce the number of uncertain quantities, and improve the accuracy of the 
theoretical predictions. Moreover, they may help us understand better the 
general features of the underlying long distance QCD effects. In the case of 
the heavy-to-heavy transitions $B \to D$ or $D^\ast$, for example, all the 
form factors are related to a single function of $q^2$, as a result of the 
Heavy Quark Symmetry (HQS) of QCD, in the limit of heavy $b$ and $c$ quarks
\cite{HQS}. 

In the case of heavy-to-light transitions, in the limit of a heavy $b$ quark, 
static in the $B$ meson rest-frame, HQS leads to the model independent 
relations \cite{IW}
\beqa
2 m_B s(q^2) &=& - f_1(q^2) 
+ \frac{m_B^2 - m_P^2}{q^2} \left[ f_0(q^2) - f_1(q^2) \right] \ , 
\label{80}\\
F_1(q^2) &=& \frac{m_B - E^\prime}{m_B + m_V} 2 V(q^2) 
+ \frac{m_B + m_V}{m_B} A_1(q^2) \ ,
\label{90} \\
F_2(q^2) &=& \frac{2 m_B |\vec{p^\prime}|^2}{(m_B + m_V)(m_B^2 - m_V^2)} 
V(q^2) + \frac{m_B - E^\prime}{m_B - m_V}  A_1(q^2) \ ,
\label{100} \\
F_3(q^2) &=& \frac{m_B E^\prime + m_V^2}{m_B (m_B + m_V)} V(q^2) 
 - \frac{m_V}{m_B} A_3(q^2) \nonumber\\
& & - \frac{m_B^2 - m_V^2}{q^2} \frac{m_V}{m_B} [A_3(q^2) - A_0(q^2)] \ .
\label{110}
\eeqa
Note that these relations are valid in any reference frame; the energy 
$E^\prime$ and momentum $|\vec{p^\prime}|$ of the light recoiling meson, $X = 
P$ or $V$, in the rest frame of the $B$ meson, are used as an abbreviation for
the more cumbersome invariant functions of $q^2$:
\beqa
E^\prime &=& \frac{m_B^2 + m_X^2 - q^2}{2 m_B} \ , \\  
\label{120}
|\vec{p^\prime}| &=& \frac{\sqrt{(m_B^2 + m_X^2 - q^2)^2  
- 4 m_B^2 m_X^2}}{2 m_B}  \ .
\label{130}  
\eeqa
Unfortunately, not much more can be said about the form factors, that relies 
solely on the properties of QCD \cite{H2L}. 

On the other hand, if one adopts the naive description of the hadronic 
transition that is provided by the Quark Model, additional relations exist 
between the form factors \cite{Stech,JMS96b,JMS98}:
\begin{itemize}
\item In Ref.~\cite{JMS98}, it was shown that, in the Quark Model, the $B \to 
V$ form factors $F_{1,2,3}$ are not independent form factors; instead, one has
\beqa 
F_1(q^2) &=& 2 A_0(q^2) \ ,  
\label{140}\\ 
\nonumber\\ 
m_V (m_B - m_V) F_2(q^2) &=& (m_B E^\prime - m_V^2) A_1(q^2) 
\nonumber\\ 
& & - \frac{2 m_B^2 |\vec{p^\prime}|^2}{(m_B + m_V)^2} A_2(q^2) \ , 
\nonumber\\ 
\label{150}\\ 
(m_B E^\prime + m_V^2) F_2(q^2)
&-& \frac{2 m_B^2 |\vec{p^\prime}|^2}{m_B^2 - m_V^2} F_3(q^2) 
\nonumber \\ 
&=& m_V (m_B + m_V) A_1(q^2) \ .
\label{160}  
\eeqa
\end{itemize}
These relations are valid for any value of the meson masses. In the special 
case of heavy-to-light transitions, that concerns us here, further relations 
can be obtained:
\begin{itemize}
\item For a heavy $b$ quark, static in the $B$ meson rest frame ($m_b \gg 
|\vec{p}_b|$), \cite{JMS96b}
\beqa
2 m_B s(q^2) &=& - f_1(q^2) + \frac{m_B^2 - m_P^2}{q^2}
\left[ f_0(q^2) - f_1(q^2) \right] \ ,
\label{170} \\
V(q^2) &=&  \frac{(m_B + m_V)^3}{2m_B m_V(E^\prime+m_V)} A_1(q^2)  
- \frac{m_B}{m_V}  A_2(q^2) \ ,
\label{180} \\
A_0(q^2) &=&  \frac{m_B + m_V}{2 m_V}
\left[ -1 + \frac{(m_B + m_V)^2}{m_B (E^\prime+m_V)} \right]  A_1(q^2)  
\nonumber\\
& & - \frac{m_B(m_B-E^\prime)}{m_V(m_B + m_V)}  A_2(q^2) \ .
\label{190}
\eeqa
\end{itemize}
It is easy to see that, together with the relations in Eqs.~\ref{140}--\ref
{160}, these Quark Model relations reproduce the model independent results 
of Eqs.~\ref{80}--\ref{110} (and provide two additional relations between $V$,
$A_{0,1,2}$).
\begin{itemize}
\item For a light $q$ quark, ultra-relativistic in the $B$ meson rest frame 
($m_q \ll |\vec{p}_q|$), \cite{JMS96b}
\beqa
f_0(q^2) &=& \left( 1 - \frac{q^2}{m_B^2 - m_P^2}
\frac{m_B+E^\prime-|\vec{p^\prime}|}
{m_B-E^\prime+|\vec{p^\prime}|} \right) f_1(q^2) \ ,
\label{200} \\
A_1(q^2) &=& \frac{2m_B|\vec{p^\prime}|}{(m_B + m_V)^2}  V(q^2)  
\label{210} \\
&=& \frac{2 m_B |\vec{p^\prime}|}{(m_B + m_V) (m_B E^\prime - m_V^2)}
\nonumber \\
& & \times \left[ \frac{m_B |\vec{p^\prime}|}{m_B + m_V} A_2(q^2)
+ m_V A_0(q^2) \right] \  .
\label{220}
\eeqa

\item Finally, in the heavy-to-light case, when both $m_b \gg |\vec{p}_b|$ and
$m_q \ll |\vec{p}_q|$ limits are considered, Eqs.~\ref{170}--\ref{190} and 
Eqs.~\ref{200}--\ref{220}, lead to five independent relations:
\beqa
f_0(q^2) &=& \left( 1 - \frac{q^2}{m_B^2 - m_P^2}
\frac{m_B+E^\prime-|\vec{p^\prime}|}
{m_B-E^\prime+|\vec{p^\prime}|} \right) f_1(q^2) \ ,
\label{230} \\
s(q^2) &=& - \frac{1}{m_B-E^\prime+|\vec{p^\prime}|} f_1(q^2) \ ,
\label{240} \\
A_1(q^2) &=& \frac{2m_B|\vec{p^\prime}|}{(m_B + m_V)^2}  V(q^2)  \ ,
\label{250} \\
A_2(q^2) &=& 
\frac{(m_B+m_V)|\vec{p^\prime}| - m_V(E^\prime + m_V)}
{m_B(E^\prime + m_V)} V(q^2) \ ,
\label{260} \\
A_0(q^2) &=& \frac{m_B-E^\prime+|\vec{p^\prime}|}{m_B + m_V} V(q^2)  \ .
\label{270}
\eeqa
\end{itemize}
With the expressions for $F_{1,2,3}$, in Eqs.~\ref{140}--\ref{160}, we obtain 
the additional relations
\beqa
F_1(q^2) &=& \frac{m_B - E^\prime + |\vec{p^\prime}|}{m_B + m_V} 2 V(q^2)  \ ,
\label{272} \\
F_2(q^2) &=& \frac{m_B - E^\prime + |\vec{p^\prime}|}{m_B - m_V} A_1(q^2) \ ,
\label{274} \\
F_3(q^2) &=& \left[ \frac{m_B + E^\prime - |\vec{p^\prime}|}{m_B + m_V} \right.
\nonumber \\
& & \left. - \left( 1 - \frac{m_V}{m_B} \right)
\frac{m_V + E^\prime - |\vec{p^\prime}|}{E^\prime + m_V} \right] V(q^2) \ .
\label{276}
\eeqa
This leaves us with two independent form factors, one for the case of a $B \to
P$ transition and one for a $B \to V$ transition.

Although these results rely on the naive Quark Model picture, where mesons are 
viewed as simple $q \overline{q}$ bound states, they do not depend on the 
particular choice for the internal momentum wavefunction of the mesons, and so
they are very general results of the Quark Model \cite{Stech,JMS96b,JMS98}. 
Our main concern in this paper is to probe to what extent these form factor 
relations remain a good approximation, beyond the Quark Model. To do so, we 
will test them against both experimental and theoretical results.

We have already pointed out that the model independent results of Eqs.~\ref{80}
--\ref{110}, which follow from the HQS of QCD in the heavy $b$ quark limit, 
are reproduced correctly by the Quark Model relations in that same limit. This 
provides a first test of the Quark Model derivations \cite{H2H}. In section 2,
we test the Quark Model relations obtained in the light $q$ quark limit. We 
explore the simple but remarkable prediction that, in that limit and due to 
the V--A structure of the weak interaction, the $\lambda = +1$ helicity 
amplitude in a $B \to V$ transition must be strongly suppressed. This effect
has been confirmed experimentally by the full angular analysis of the $B \to 
K^\ast J/\psi$ decay, performed by the CLEO Collaboration. We show that the 
theoretical expression for the $\lambda = +1$ amplitude, written in terms of 
the usual hadronic form factors, is not suppressed, unless the form factors
relations are verified. A further test is provided by the ratio of the two
remaining helicity amplitudes, with $\lambda = 0$ and $-1$; using the Quark 
Model relations, that ratio can be predicted and it is in excellent agreement
with the CLEO data for the $B \to K^\ast J/\psi$ decay. We also discuss 
analogous tests that can be carried out for semileptonic decays, such as $B 
\to \rho l^- \overline{\nu}_l$, where strong interaction effects are less of a 
problem and a range of recoil momenta is available. In section 3, the Quark 
Model form factor relations are compared to Lattice QCD and Light Cone 
Sum-Rules predictions --- two methods that provide model independent, albeight
approximate, insights into the long distance QCD effects that are at play in 
the hadronic transitions. Here too the agreement with the Quark Model results 
is striking.

\section{Helicity amplitudes in heavy-to-light\\ transitions}

\subsection{$B \to K^\ast J/\psi$}

\subsubsection{Helicity $\lambda = +1$ amplitude}

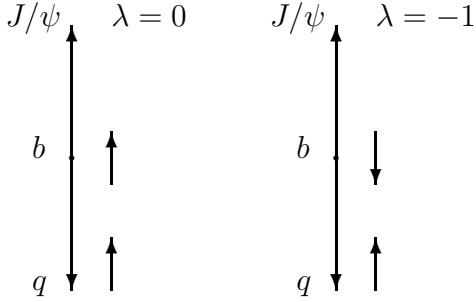
\begin{figure}[ht]
\centering
\begin{picture}(200,200)(0,0)
\thicklines
\put(50,100){\circle*{2}}
\put(50,100){\vector(0,-1){50}}
\put(50,100){\vector(0,1){50}}
\put(150,100){\circle*{2}}
\put(150,100){\vector(0,-1){50}}
\put(150,100){\vector(0,1){50}}
\put(65,90){\vector(0,1){20}}
\put(65,50){\vector(0,1){20}}
\put(65,150){$\lambda=0$}
\put(35,100){$b$}
\put(35,50){$q$}
\put(25,150){$J/\psi$}
\put(165,110){\vector(0,-1){20}}
\put(165,50){\vector(0,1){20}}
\put(165,150){$\lambda=-1$}
\put(135,100){$b$}
\put(135,50){$q$}
\put(125,150){$J/\psi$}
\end{picture}
\caption{Allowed helicity amplitudes in $b \to q + J/\psi$}
\label{fig1}
\end{figure}

Let us consider heavy-to-light meson transitions where the underlying quark 
process is the weak decay of a heavy $b$ quark into a light quark $q=u$, $d$
or $s$ and a spin-1 particle. One such process is the inclusive $J/\psi$
production mechanism
\beq
b \to q + (c \overline{c})_{J/\psi} \ ,
\label{300}
\eeq
where $q = s$ or $d$ and the $c \overline{c}$ pair hadronizes into a $J/\psi$.
The weak charged current at the origin of the decay produces a light $q$ quark
that is left-handed. In the limit $m_q \to 0$ (or, more precisely, in the
ultra-relativistic limit $m_q \ll |\vec{p}_q|$), the left-handed light quark 
has negative helicity and so, from angular momentum conservation, the spin-1 
state $J/\psi$ can have helicity $\lambda = 0$ or $-1$, but not $\lambda = +1$.
The diagrams in Fig.~\ref{fig1} illustrate this simple but remarkable 
consequence of the V--A structure of the charged weak interaction.

We can check that an explicit calculation of the $b \to q + J/\psi$ decay rate,
for the different helicity states of the $J/\psi$, leads to the same 
conclusion. 
The decay amplitude can be written in the form of an effective $bqJ/\psi$ weak 
vertex. where one must be careful to take into account the effects of the 
strong interaction. In particular, the exchange of gluons can generate terms 
with Lorentz structures that are not present in the weak vertex, in the 
absence of QCD. For that reason, we write the effective $bqJ/\psi$ weak vertex
in its most general form \cite{JMS96a},
\beqa
\Lambda^\mu_{bqJ/\psi} &=& 
- \frac{G_F}{\sqrt{2}} \: V_{cb}V_{cq}^\ast \: \frac{f_{J/\psi}}{m_{J/\psi}}
\overline{q} \left\{ G_0 \: k^\mu \not{\!k} (1 - \gamma_5) \right.  
\nonumber\\ 
& & + G_1 \: (k^2 g^{\mu\nu} - k^\mu k^\nu) \gamma_\nu (1 - \gamma_5) 
\nonumber\\
& & \left. + G_2 \: i \sigma_{\mu\nu} k^\nu 
\left[ m_b (1 + \gamma_5) + m_q (1 - \gamma_5) \right] \right\} b  \ ,
\label{310}
\eeqa
where $k \equiv p_b - p_q = p_{J/\psi}$. Each term in the vertex is multiplied 
by a coefficient $G_i$ ($i = 0,1,2$), that includes both perturbative and 
non-perturbative QCD effects, but whose value is not of immediate importance 
for our discussion. Notice that the $G_0$ term, proportional to $k^\mu$, does 
not contribute to the decay amplitude, and it is only included for 
completeness. Notice also the important fact that only left-handed quark fields
participate in the interaction (in the $G_2$ term, we have used the equations 
of motion to write that term in a more familiar form). The inclusive $b \to q 
+ J/\psi$ decay rate that follows is
\beqa
\Gamma_\lambda &=& \frac{1}{4 \pi} G_F^2 |V_{cb}V_{cq}^\ast|^2
\frac{|\vec{p}_q|}{m_b} \frac{f_{J/\psi}^2}{m_{J/\psi}^2}
\nonumber \\
& & \times \left\{ \begin{array}{lll}
m_{J/\psi}^2 \left| G_1 - G_2 \right|^2 
\left[ (m_b^2 + m_q^2) E_q - 2 m_b m_q^2 \right] 
& & \lambda = 0 \\
& & \\
( E_q \mp |\vec{p}_q| ) \left| G_1 m_{J/\psi}^2 
- G_2 ( m_{J/\psi}^2\mp 2 m_b |\vec{p}_q| ) \right|^2 
& & \lambda = \pm 1 
\end{array} \right.
\label{320}
\eeqa
where $E_q$ and $|\vec{p}_q|$ are the energy and momentum of the $q$ quark in 
the $b$ rest-frame. We see that the decay rate for a $J/\psi$ with helicity 
$\lambda = +1$ does indeed vanish, in the limit $m_q \to 0$. 

The conclusion that the $\lambda = +1$ decay rate is strongly suppressed must 
also apply to each one of the exclusive decay channels that add up to the 
inclusive $b \to q + J/\psi$ process. In particular, when the light quark $q 
= s, d$ hadronizes into a vector meson $V = K^\ast, \rho, \omega$ or $\phi$ 
(for a $B_s$ decay), the two body decay $B \to V J/\psi$ could {\it a priori}
have three helicity final states $\lambda = 0, \pm 1$ (the two vector mesons 
must have the same helicity $\lambda$); but the argument above tells us that 
the $\lambda = +1$ amplitude is very suppressed, and vanishes in the limit 
$m_q \to 0$.

Contrary to the case of the inclusive process, the suppression of the $\lambda
= +1$ amplitude in the exclusive decay $B \to V J/\psi$ can be tested 
experimentally: by analyzing the angular distribution of the decay products
of the $J/\psi$ and of the vector meson $V$, one can determine the relative
sizes and phases of all three helicity amplitudes. This has been done
recently for the decay
\beq
\begin{picture}(50,50)(0,0)
\put(10,35){$B \to K^\ast \; J/\psi$ \ ,}
\put(45,30){\line(0,-1){25}}
\put(45,5){\vector(1,0){30}}
\put(80,0){$K \pi$}
\put(70,30){\line(0,-1){10}}
\put(70,20){\vector(1,0){30}}
\put(105,15){$l^+ l^-$}
\end{picture}
\label{330}
\eeq
by the CLEO Collaboration \cite{CLEO}. The CLEO analysis was able to determine
the differential decay rate
\beqa
\lefteqn{
\frac{1}{\Gamma} \frac{d^3\Gamma}{d\cos\theta_{K^\ast} d\cos\theta_\psi 
d\chi} = \frac{9}{16\pi} \left\{ 
\sin^2\theta_\psi \cos^2\theta_{K^\ast} |H_0|^2 \right.} 
\nonumber \\
& & + \frac{1}{4} (1 + \cos^2\theta_\psi)
\sin^2\theta_{K^\ast} (|H_+|^2 + |H_-|^2) 
\nonumber \\
& & - \frac{1}{2} \sin^2\theta_\psi \sin^2\theta_{K^\ast}
\left[ \cos2\chi Re(H_+H_-^\ast) - \sin2\chi Im(H_+H_-^\ast) \right]
\nonumber\\
& & - \frac{1}{4} \sin2\theta_\psi \sin2\theta_{K^\ast}
\left[ \cos\chi Re(H_+H_0^\ast+H_-H_0^\ast) \right.
\nonumber \\
& & \left. \left. - \sin\chi Im(H_+H_0^\ast-H_-H_0^\ast) \right] \right\} \ ,
\label{340}
\eeqa
where $\theta_\psi$ [resp. $\theta_{K^\ast}$] is the angle between the 
$l^+$ [resp. $K$] and the $J/\psi$ [resp. $K^\ast$] momenta, in the rest-frame
of the vector meson, and $\chi$ is the asymuthal angle between the decay planes
of the $J/\psi$ and the $K^\ast$. The coefficients $H_{0,\pm}$ are proportional
to the helicity amplitudes $A_{0,\pm}$ for the $B \to K^\ast J/\psi$ decay:
\beq
H_i = \frac{A_i}{|A_+|^2 + |A_-|^2 + |A_0|^2} \;\;\;\; (i = 0, \pm) \ .
\label{350}
\eeq
The experimental analysis of Ref.~\cite{CLEO} determined the magnitude and 
phase of the ratios
\beq
r_\pm \equiv \frac{H_\pm}{H_0} = \frac{A_\pm}{A_0} \ ;
\label{360}
\eeq
they are (after translating the experimental results from the transversity
basis to the helicity basis used in here)
\beqa
|r_+|^2 &=& 0.03 \pm 0.07 \label{362} \\
|r_-|^2 &=& 0.90 \pm 0.28 \label{364} \\
Arg(r_+) &=& 2.9 \pm 1.7 \label{366} \\
Arg(r_-) &=& 3.01 \pm 0.29 \ . \label{368}
\eeqa
The size of the $\lambda = +1$ amplitude, $A_+$, is consistent with zero, 
and much smaller than the other two helicity amplitudes --- this is the effect 
that was predicted above. The relative phases $\phi_{\pm}$ between $A_\pm$ 
and $A_0$ are consistent with $\pi$ (the large error in the phase of $r_+$ is 
due to the small size of the $\lambda = +1$ amplitude), and show no signs of 
imaginary parts in the decay amplitudes. This is important for our discussion, 
as the presence of imaginary terms would signal the existence of significant 
final state interaction effects. It would then be possible for the $J/\psi$ to
scatter from one helicity state to another invalidating, or at least weakening,
our argument for the suppression of the $\lambda = +1$ amplitude. Later on, we
will consider other decays where this potential problem is not a concern.

With the simple argument for the suppression of the $\lambda = +1$ amplitude 
in $B \to V J/\psi$ confirmed by experiment, we want to see how that 
suppression appears explicitly in the theoretical expressions for the helicity
amplitudes, $A_\lambda$. These amplitudes are obtained from the matrix element
of the effective $bqJ/\psi$ vertex of Eq.~\ref{310}, between the $B$ and $V$ 
meson states; we use the form factors of Section 1 to parametrize the different
terms in the expression. The results are
\beqa
A_0 &=& 
- \frac{G_F}{\sqrt{2}} \: V_{cb}V_{cq}^\ast \: 
\frac{f_{J/\psi}}{m_{J/\psi}} \frac{m_{J/\psi}}{2 m_V}
\nonumber\\
& & \times \left\{ G_1 \: \left[ (m_B + m_V) (m_B^2 - m_V^2 - m_{J/\psi}^2) A_1
- \frac{4 m_B^2 |\vec{p_V}|^2}{m_B + m_V} A_2 \right] \right.
\nonumber\\
& & \left. + G_2 (m_b - m_q) 
\: \left[ - (m_B^2 + 3 m_V^2 - m_{J/\psi}^2) F_2
+ \frac{4 m_B^2 |\vec{p_V}|^2}{m_B^2 - m_V^2} F_3 \right] \right\}
\nonumber \\
& &
\label{380}
\eeqa
for the longitudinal amplitude, and
\beqa
A_\pm &=& 
- \frac{G_F}{\sqrt{2}} \: V_{cb}V_{cq}^\ast \: \frac{f_{J/\psi}}{m_{J/\psi}}
\nonumber\\
& & \times \left\{ G_1 m_{J/\psi}^2 \: \left[ - (m_B + m_V) A_1 \pm
\frac{2 m_B |\vec{p_V}|}{m_B + m_V} V \right] \right.
\nonumber\\
& & + G_2 m_b \: \left[ \mp  m_B |\vec{p_V}| F_1 
+ (m_B^2 - m_V^2) F_2 \right]
\nonumber\\
& & \left. + G_2 m_q \: \left[ \mp  m_B |\vec{p_V}| F_1 
- (m_B^2 - m_V^2) F_2 \right] \right\}
\label{390}
\eeqa
for the transversal amplitudes; the form factors are evaluated at $q^2 = 
m_{J/\psi}^2$. It is easy to check that the $\lambda = +1$ amplitude will 
vanish, in the limit $m_q \to 0$, {\it provided the form factors obey precisely
the Quark Model relations} of Eqs.~\ref{200}-\ref{220} (together with the 
relations in Eqs.~\ref{140}-\ref{160}) that are valid in that same limit 
\cite{Comment}. Without those relations, the form factors that appear in the 
helicity amplitudes are independent of each other, and the large suppression 
in $A_+$, that is expected and which has been confirmed experimentally in the 
case of $B \to K^\ast J/\psi$, cannot be accounted for.

\subsubsection{Helicity $\lambda = -1$ amplitude}

If, in addition to the limit of an ultra-relativistic light $q$ quark ($m_q 
\ll |\vec{p}_q|$), we adopt the limit of a static heavy $b$ quark ($m_b 
\gg |\vec{p}_b|$), the set of Quark Model relations is extended to those in 
Eqs.~\ref{230}-\ref{276}. It is then possible to predict, in addition to 
$A_+/A_0 = 0$, the value of the ratio $A_-/A_0$. In fact, applying the 
additional form factor relations to the amplitudes in Eqs.~\ref{380} and
\ref{390}, we can write $A_0$ and $A_-$ in terms of a single form factor;
the dependence on that form factor can then be eliminated by considering the
ratio
\beqa
\frac{A_-}{A_0} &=& - \frac{2 m_{J/\psi}}{m_B - E_V + |\vec{p_V}|}
\frac{1  - \frac{G_2}{G_1} m_b 
(m_B - E_V + |\vec{p_V}|)/m_{J/\psi}^2}
{1 - \frac{G_2}{G_1} m_b/(m_B - E_V + |\vec{p_V}|)} \ .
\label{420}
\eeqa
In order to compare this prediction to the experimental result from the CLEO 
analysis, we must first determine the ratio $G_2/G_1$ of the parameters that 
appear in the effective $bqJ/\psi$ vertex of Eq.~\ref{310}. 

If not for the QCD corrections to the weak $bqJ/\psi$ vertex, $G_2/G_1 = 0$. On
the other hand, if the predominant QCD corrections were short-distance in 
nature, this ratio could be calculated perturbatively from the Feynman 
diagrams for the vertex, with the appropriate gluon exchanges \cite{BE}.
However, the large discrepancy between the perturbative calculation for the 
inclusive $b \to q + J/\psi$ decay rate and the experimental result tells us 
that large non-perturbative QCD effects are at play, and so a theoretical 
estimate of $G_2/G_1$ becomes very hard to obtain. Instead, we can derive this
ratio from the measurement of the $J/\psi$ polarization, $P \equiv 
\Gamma_L/\Gamma$, in the inclusive decay. From Eq.~\ref{320}, with $m_q = 0$, 
we obtain
\beqa
P &=& \left[ 1 + \frac{2|m_{J/\psi}^2/m_b^2 - G_2/G_1|^2}
{m_{J/\psi}^2/m_b^2 |1-G_2/G_1|^2} \right]^{-1} \ .
\label{430}
\eeqa
On the other hand, the experimental result for the polarization is 
\cite{CLEO95}
\beqa
P^\prime &=& 0.59 \pm 0.15 \ ,
\label{440}
\eeqa
where the prime reminds us that the measurement is contaminated by the 
contribution from decays of $B$ mesons into higher charmonium states, that in 
turn decay to $J/\psi$. These cascade decays account for the substantial 
difference between the inclusive $B \to J/\psi + X$ branching ratio $B^\prime 
= (1.13 \pm 0.06) \%$ \cite{PDG}, and the branching ratio for the direct decay
$b \to q + J/\psi$, $B = (0.80 \pm 0.08)\%$ \cite{PDG}. In order to obtain a 
value for the polarization of the $J/\psi$ in the direct decay, we assume (to 
lack of a better estimate) that the cascade decays produce unpolarized 
$J/\psi$s. Then, the experimental result of Eq.~\ref{440} translates into
\beq
P = 0.70 \pm 0.22 \ .
\label{450}
\eeq
Comparing with the theoretical prediction of Eq.~\ref{430}, and taking $m_b = 
5.0$ GeV, we obtain
\beq
\frac{G_2}{G_1} = 0.13 \pm 0.18 \ .
\label{460}
\eeq
(There is a second, larger, solution for $G_2/G_1$ that can be discarded, as 
discussed in Ref.~\cite{JMS96a}). Notice that the error is very large, and does
not include the uncertainty in the polarization of the $J/\psi$s from the 
cascade decays. 

Applying this estimate of $G_2/G_1$ to Eq.~\ref{420}, we obtain the following
prediction for the ratio of the $\lambda = 0$ and $-1$ helicity amplitudes in 
$B \to K^\ast J/\psi$:
\beq
\frac{A_-}{A_0} = - 0.93 \pm 0.48 \ .
\label{470}
\eeq
This is to be compared with the experimental results for the magnitude and 
phase of $r_-$, in Eqs.~\ref{364}-\ref{368}. Despite the large error in the 
theoretical prediction, it is clear that it gives the correct relative sign 
between the two amplitudes. The central value in the theoretical estimate for 
the magnitude of $A_-/A_0$ is also in excellent agreement with the 
experimental result, $|r_-| = 0.95 \pm 0.15$.

\subsection{$B \to \rho l^- \overline{\nu}_l$}

Similar tests of the Quark Model relations can be performed with semileptonic 
decays of the type
\beq
\begin{picture}(35,35)(0,0)
\put(10,20){$b \to u + W^\ast$ \ ,}
\put(60,15){\line(0,-1){10}}
\put(60,5){\vector(1,0){30}}
\put(95,0){$l^- \overline{\nu}_l$}
\end{picture}
\label{480}
\eeq
with the light $u$ quark hadronizing into a vector meson $V = \rho, \omega$ 
or $K^\ast$ (for a $B_s$ decay). As before, the V--A structure of the charged 
weak interaction produces a left-handed $u$ quark; in the limit $m_u \ll
|\vec{p}_u|$, the ultra-relativistic quark has negative helicity and so the 
virtual $W$ cannot have helicity $\lambda = +1$. In the case of the exclusive 
decays, where both the vector meson and the virtual $W$ have the same helicity,
this translates into the cancellation of the $\lambda = +1$ amplitude, in the
limit $m_u \to 0$.

As with the case of the hadronic decay, the effect can be seen explicitly in 
the theoretical expression for the inclusive semi-leptonic decay rate. For $b 
\to u + W^{\ast} \to u + l^- \overline{\nu}_l$, the differential decay rate is 
\beqa
\frac{d\Gamma_\lambda}{dk^2/m_b^2} &=& \frac{1}{24 \pi^3} G_F^2 |V_{ub}|^2
m_b |\vec{p}_u| \nonumber \\
& & \times \left\{ \begin{array}{lll}
2 m_b |\vec{p}_u|^2 + E_u k^2 & & \lambda = 0 \\
& & \\
(E_u \mp |\vec{p}_u|) k^2 & & \lambda = \pm 1 
\end{array} \right. \ ,
\label{500}
\eeqa
where $k \equiv p_b - p_u$, and $E_u$ and $|\vec{p}_u|$ are the energy and 
momentum of the $u$ quark, in the $b$ rest-frame. As expected, the decay rate 
for the $\lambda = +1$ helicity of the virtual $W$ vanishes, in the limit $m_u 
\to 0$. On the other hand, for the exclusive semi-leptonic decays $B \to V + 
W^\ast \to V + l^- \overline{\nu}_l$, with the $B \to V$ hadronic matrix 
element parametrized by the form factors of Section 1, the differential decay 
rate is
\beqa
\frac{d\Gamma_\lambda}{dq^2/m_B^2} &=& 
\frac{1}{96 \pi^3} G_F^2 |V_{ub}|^2 |\vec{p}_V| q^2 |H_\lambda(q^2)|^2 \ ,
\label{520}
\eeqa
where
\beqa
2 m_V \sqrt{q^2} H_0(q^2) &=&
(m_B + m_V) (m_B^2 - m_V^2 - q^2) A_1(q^2)
\nonumber \\
& & - \frac{4 m_B^2 |\vec{p}_V|^2}{m_B + m_V} A_2(q^2) \ ,
\label{530} \\
H_\pm(q^2) &=& \frac{2 m_B |\vec{p}_V|}{m_B + m_V} V(q^2)
\mp (m_B + m_V) A_1(q^2)
\label{540}
\eeqa
and $q \equiv p_B - p_V$.
As in the case of the hadronic decay, we find that we need the form factor 
relations of Eqs.~\ref{200}-\ref{220} in order for $H_{+1}$ to vanish, when 
$m_u \to 0$. 

In the limit of a static heavy $b$ quark and an ultra-relativistic light $u$ 
quark, in addition to $|H_+|^2/|H_0|^2 = 0$, the extended form factor 
relations of Eqs.~\ref{230}-\ref{276} lead to the prediction that
\beq
\frac{|H_-|^2}{|H_0|^2} = \frac{4 q^2}{(m_B - E_V + |\vec{p}_V|)^2} \ .
\label{550}
\eeq
This provides another possible test of the form factor relations, in analogy to
the case of the hadronic $B \to V J/\psi$ decay.

At present, there is no experimental data to compare these predictions to. In
fact, using the semileptonic decay $B \to \rho l^- \overline{\nu}_l$ to test 
the Quark Model form factor relations, instead of the $B \to K^\ast J/\psi$ 
decay, requires a substantial experimental effort, due to the additional 
Cabibbo suppression. There are however significant advantages to using the 
semileptonic decay. The first 
one is that no strong final state interactions are present that could scatter 
the vector meson between states of different helicities. This makes the 
argument for the suppression of the $\lambda = +1$ amplitude much stronger 
than in the case of an hadronic decay. The other advantage is that the 
magnitude of the vector meson momentum is not fixed in the three body decay. 
This allows testing the form factor relations throughout the entire range of 
$q^2$, from $q^2 =0$ (maximum recoil) to $q^2 = q^2_{max} = (m_B - m_V)^2$
(zero recoil).

\subsection{$B \to K^\ast \gamma$}

Other well known decays, of the same type as those discussed in here, are the 
inclusive radiative decay $b \to q + \gamma$, with $q = s$ or $d$, and the 
corresponding exclusive decays $B \to V \gamma$, with $V = K^\ast, \rho, 
\omega$ or $\phi$ (for a $B_s$ decay). Unfortunately, this special 
case of the $B \to V$ transition, with $q^2 = (p_B - p_V)^2 = 0$, cannot be 
used to test the form factor relations.

The discussion for the radiative decay is analogous to that for the hadronic
decay into $J/\psi$. The $bq\gamma$ vertex is similar to the $bqJ/\psi$ vertex
of Eq.~\ref{310}, with the CKM factor $|V_{tb}V_{tq}^\ast|$ replacing 
$|V_{cb}V_{cq}^\ast|$ and the $G_0$ term omitted, to preserve gauge invariance:
\beqa
\Lambda^\mu_{bq\gamma} &=& 
- \frac{G_F}{\sqrt{2}} \: V_{tb}V_{tq}^\ast \: 
\overline{q} \left\{ G^\prime_1 \: (k^2 g^{\mu\nu} - k^\mu k^\nu) \gamma_\nu 
(1 - \gamma_5) \right.
\nonumber\\
& & \left. + G^\prime_2 \: i \sigma_{\mu\nu} k^\nu 
\right[ m_b (1 + \gamma_5) + m_q (1 - \gamma_5) \left] \right\} b  \ .
\label{560}
\eeqa
Using $k^2 = 0$ to simplify our expressions, the decay rate that follows is
\beqa
\Gamma_\lambda &=& \frac{1}{8 \pi} G_F^2 |V_{tb}V_{tq}^\ast|^2
\left( 1 - \frac{m^2_q}{m^2_b} \right) |G^\prime_2|^2 m_b^3
\; \left\{ \begin{array}{lll}
m_b^2 & & \lambda = - 1 \\ 
m_q^2 & & \lambda = + 1
\end{array} \right. \ .
\label{570}
\eeqa
Again, the $\lambda = +1$ rate is very suppressed and vanishes in the limit 
$m_q \to 0$. However, and contrary to the more general case of the $b \to q + 
J/\psi$ decay, that is a trivial result, since the only contribution to the 
$\lambda = +1$ rate comes from the term proportional to $G^\prime_2 m_q$, in 
the effective vertex. The same conclusion applies to the decay amplitude 
for the exclusive decay $B \to V \gamma$,
\beqa
A_\pm = - \frac{G_F}{\sqrt{2}} \: V_{tb}V_{tq}^\ast \:
(m_B^2 - m_V^2) G^\prime_2 2 F_2(0) 
\; \left\{ \begin{array}{lll}
m_b & & \lambda = - 1 \\ 
(-) m_q & & \lambda = + 1
\end{array} \right. \ .
\label{580}
\eeqa
The $\lambda = +1$ 
amplitude only has contributions from the term proportional to $G^\prime_2 
m_q$ in the vertex, and it vanishes automatically in the limit $m_q \to 0$. 
There is no need to impose any constraint on the form factors and so the 
radiative decays cannot be used as a test of the Quark Model relations.

\section{Heavy-to-light Form Factors\\ in Lattice QCD
and Light Cone Sum-Rules}

\begin{figure}
\unitlength0.05in
\begin{picture}(100,60)

\put(0,10){\makebox(50,30)
{\psfig{figure=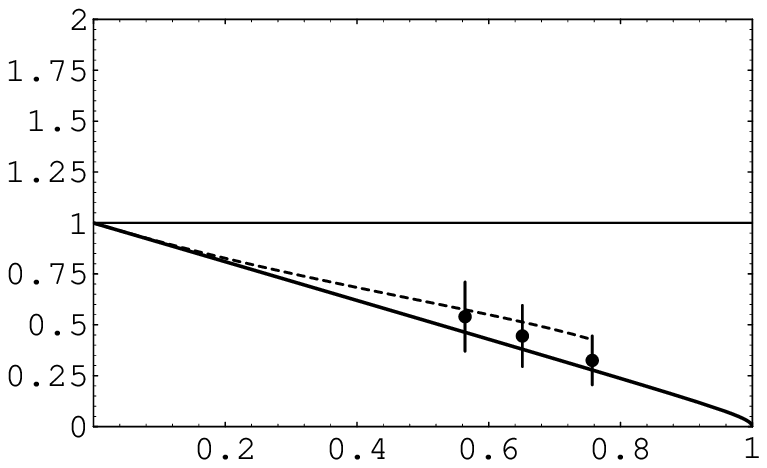,width=2.7in}}}
\put(6,35){{\large a) $f_0/f_1$}}
\put(21,3){{\large $q^2/q^2_{max}$}}

\put(55,10){\makebox(50,30)
{\psfig{figure=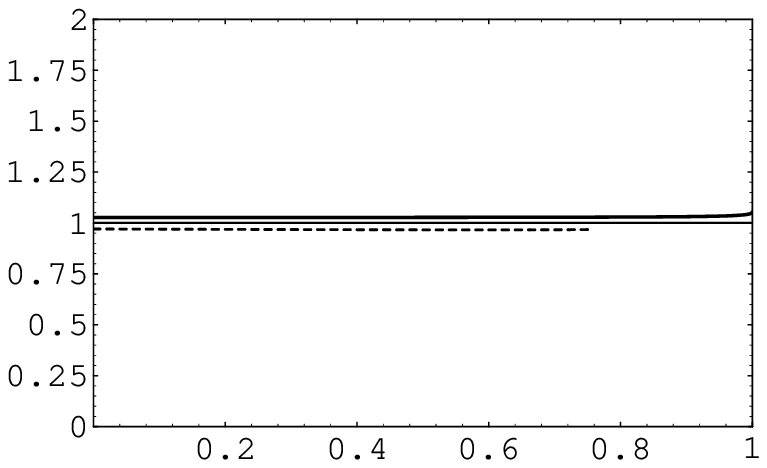,width=2.7in}}}
\put(61,35){{\large b) $-s (m_B + m_\pi)/f_1$}}
\put(76,3){{\large $q^2/q^2_{max}$}}

\end{picture}

\caption{$B \to \pi$ form factor ratios in the Quark Model (full lines),
Light Cone Sum-Rules (dashed lines) and Lattice QCD (data points).}
\label{fig2}
\end{figure}

\begin{figure}
\unitlength0.05in
\begin{picture}(100,120)

\put(0,10){\makebox(50,30)
{\epsfig{figure=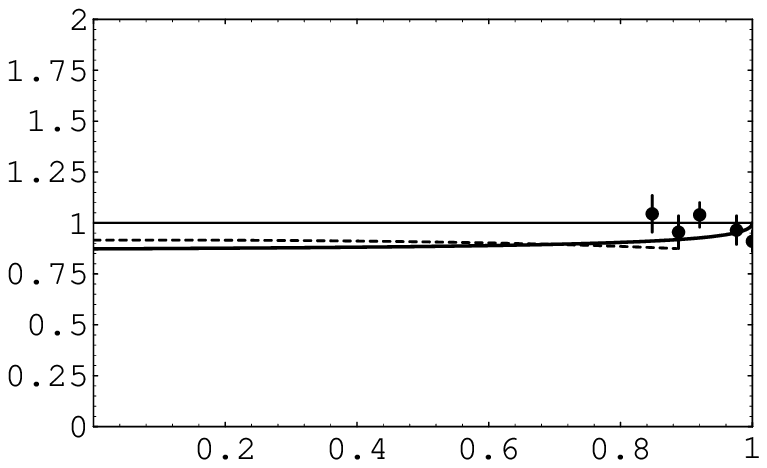,width=2.7in}}}
\put(6,35){{\large e) $A_1/F_2$}}
\put(21,3){{\large $q^2/q^2_{max}$}}

\put(55,10){\makebox(50,30)
{\epsfig{figure=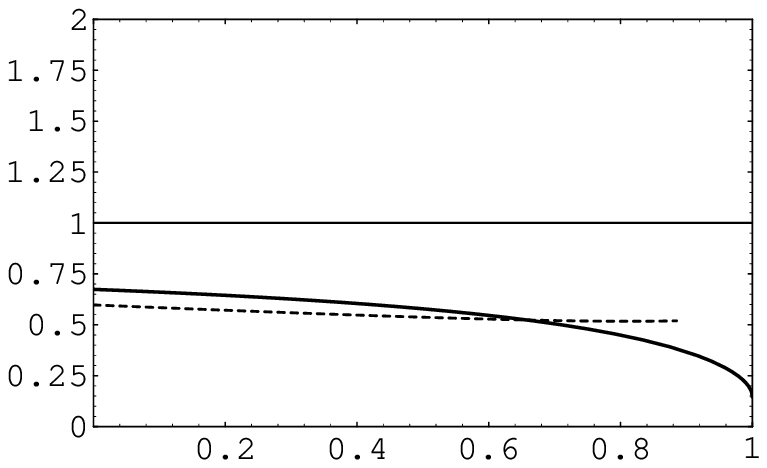,width=2.7in}}}
\put(61,35){{\large f) $F_3/V$}}
\put(76,3){{\large $q^2/q^2_{max}$}}

\put(0,45){\makebox(50,30)
{\epsfig{figure=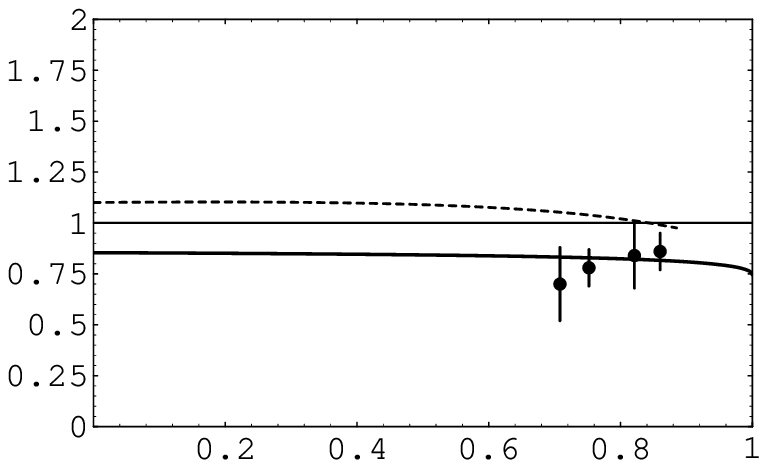,width=2.7in}}}
\put(6,70){{\large c) $A_0/V$}}

\put(55,45){\makebox(50,30)
{\epsfig{figure=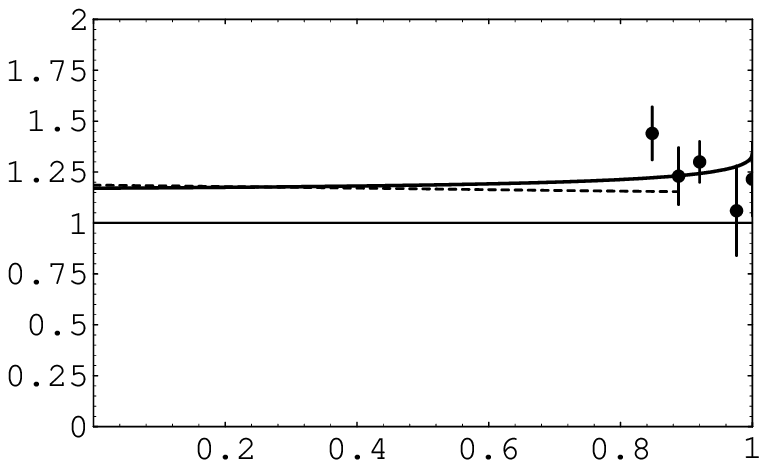,width=2.7in}}}
\put(61,70){{\large d) $2 V/F_1$}}

\put(0,80){\makebox(50,30)
{\epsfig{figure=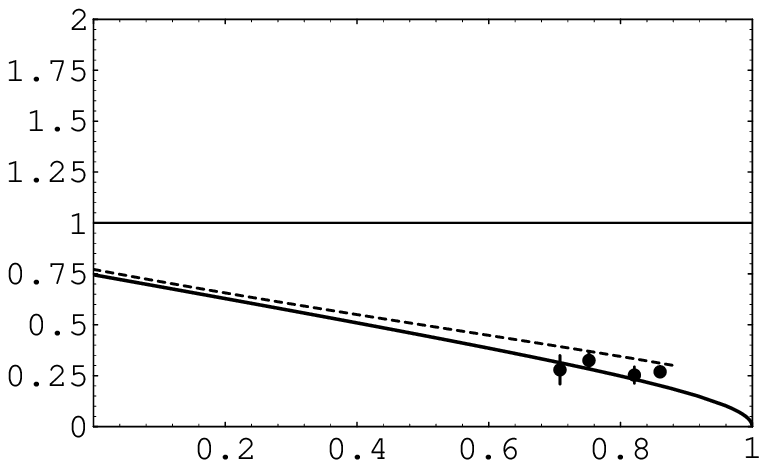,width=2.7in}}}
\put(6,105){{\large a) $A_1/V$}}

\put(55,80){\makebox(50,30)
{\epsfig{figure=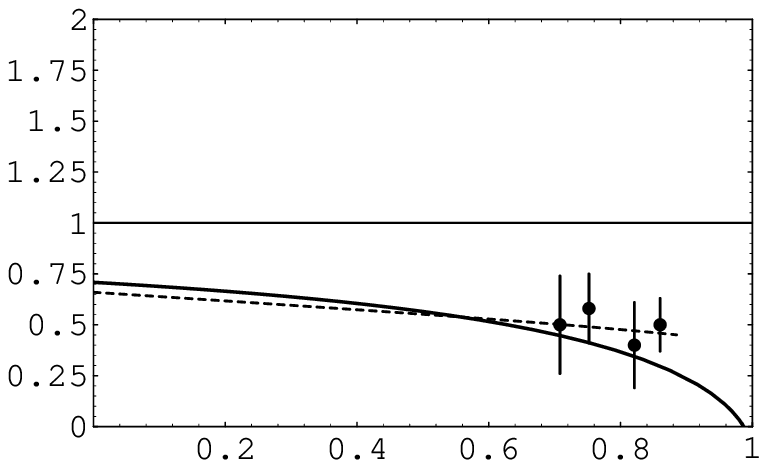,width=2.7in}}}
\put(61,105){{\large b) $A_2/V$}}

\end{picture}
\caption{$B \to \rho$ form factor ratios in the Quark Model (full lines),
Light Cone Sum-Rules (dashed lines) and Lattice QCD (data points).}
\label{fig3}
\end{figure}

A comparison to the predictions of Lattice QCD and of Light Cone Sum-Rules 
(LCSR) provides another useful test of the Quark Model form factor relations.
In Figs.~\ref{fig2} and \ref{fig3}, we plotted the ratios of form factors 
obtained from Eqs.~\ref{230}-\ref{276}, for the case of $B \to \pi$ and $B \to
\rho$ transitions. Also shown in the same figures are the results obtained 
from the recent LCSR calculations of Ref.~\cite{LCSR} and from the Lattice
results of Ref.~\cite{Lattice}. The LCSR predictions are valid in a sizable 
range of $q^2$ that only excludes the low recoil (high $q^2$) region; 
conversely, that is the region where it is possible to obtain reliable Lattice
results, for heavy-to-light transitions. The LCSR curves should be understood 
as accompanied by an error estimated at about $20\%$ \cite{LCSR}, that is not 
shown in the plots. Also not shown is the systematic error in the Lattice data
points; the error bars correspond to the statistical error only. The agreement
between the Quark Model predictions and both the Lattice QCD and the LCSR 
results is remarkable. It is worth pointing out, in particular, how such small
effects as the predicted deviations from unity, in the ratios $A_0/V$ and $2 
V/F_1$ (an effect of the order of $m_V/m_B$), are well supported by 
the Lattice results.

We have plotted the Quark Model form factor
ratios for the entire range of $q^2$. However, the results near
$q^2 = 0$ (maximum recoil) and $q^2 = q^2_{max}$ (zero recoil) must be viewed 
with care, since the static $b$ quark and the ultra-relativistic $q$ quark 
limits may not be entirely justified in these regions. 
In order to estimate at what point, and by how much, these 
approximations may fail, we would need some information about the quark 
dynamics inside the mesons (in particular, we would need to estimate the 
ratios $m_b/|\vec{p}_b|$ and $|\vec{p}_q|/m_q$ as a function of $q^2$). In 
Ref.~\cite{JMS96b}, a few simple scenarios for the internal momentum 
wavefunctions of the mesons were discussed, that led to estimates of the quark
momenta relative to their masses. These, however, are rather simplistic and
model dependent results, that can serve an illustrative purpose, but cannot 
provide a reliable estimate of the range of $q^2$ where the form factor 
relations are valid. Instead, our rationale in here is to determine that range
of validity through a comparison to experimental data or model independent 
theoretical methods, such as Lattice QCD and LCSR. Surprisingly, no significant
discrepancies can be found, at the present level of accuracy, for any of the
tests that were performed, and at any value of $q^2$.

\section{Conclusion}

We have compared the Quark Model form factor relations of Refs.~\cite{Stech,
JMS96b, JMS98}, to both experimental results and the theoretical predictions
of Lattice QCD and Light Cone Sum-Rules. In every case, the agreement is 
impressive and suggests that the relations remain a good approximation, beyond
the Quark Model. If that proves to be the case, then large theoretical 
uncertainties associated with hadronic form factors can be avoided, in the 
study of heavy-to-light weak transitions. For example, one important 
application would be obtaining a measurement of $|V_{ub}|$ that is independent
of hadronic form factors, as suggested in Refs.~\cite{JMS98} or \cite{Vub}. 
But beforehand, one should try to improve the tests
of the form factor relations that were discussed in here.
In the $B \to K^\ast J/\psi$ analysis, it would be interesting to improve the
precision in the measurement of $|A_+/A_0|$, to the 
point were the deviation from an exact cancellation of $A_+$ can be detected. 
By itself, such a measurement could not be easily interpreted:
a non-vanishing $A_+$ amplitude could be due to corrections to the Quark Model
relations, but it could also be due to corrections from a non-vanishing 
light quark mass. To help desintangle both effects, 
the analysis can be repeated for the analogous, but Cabibbo suppressed, $B \to
\rho J/\psi$ decay. There, the light quark is $q = u$ and one is likely to be 
much closer to the $m_q = 0$ limit where the Quark Model derivation applies.
Even better would be to determine the helicity amplitudes in the 
semileptonic decay $B \to \rho l^- \overline{\nu}_l$, and repeat the
test of the Quark Model relations, at different values of $q^2$, and 
without the added uncertainty of the effects of strong final state 
interactions.
Another area for improvement is in the measurement of the $J/\psi$ 
polarization in $b \to q + J/\psi$. As we have seen, together with the
measurement of the  $b \to q + J/\psi$ branching ratio, that would fully 
determine the effective
$bqJ/\psi$ vertex. From there, we could better predict the ratio $A_-/A_0$ in 
$B \to V J/\psi$, and this additional test of the Quark Model 
relations could be improved. Again, the same test can be performed
with the semileptonic decay, without any of the theoretical uncertainties 
regarding the form of the weak vertex.

\section*{Acknowledgments} 

This research was funded by a grant from the National Science Foundation.


\begin{thebibliography}{abcd}
\bibitem{H2H}
It is worth pointing out that a Quark Model derivation, analogous to that 
discussed in here, can be repeated for the case of heavy-to-heavy transitions
\cite{Stech,JMS96b}, and it leads to the model independent form factor 
relations that follow from HQS \cite{HQS}.
\bibitem{Comment}
The cancellation would not be exact, if one were to use the relations derived 
in Ref.~\cite{Stech}; that is due to the small disagreement in terms of order 
$m_V/m_B$ that was already pointed out.
\end{thebibliography}
\end{document}